\documentclass[preprint,12pt]{elsarticle}
\usepackage{}

\usepackage{amsmath}
\usepackage{cases}
\usepackage{color}
\usepackage{hyperref}
\usepackage{amssymb}
\usepackage{graphicx}
\usepackage{graphics}
\usepackage{subfigure}
\usepackage{slashbox}
\usepackage{appendix}
\usepackage{pict2e}
\usepackage{textcomp}
\usepackage{bm}
\usepackage{url}

\makeatletter
  \newcommand\figcaption{\def\@captype{figure}\caption}
  \newcommand\tabcaption{\def\@captype{table}\caption}
\makeatother

\biboptions{numbers,sort&compress}

\begin{document}

\begin{center}

{\Large Observation of  two coupled  Faraday  waves  in a vertically vibrating Hele-Shaw cell  with  one of them oscillating  horizontally}

\vspace{0.3cm}

Xiaochen Li, Xiaoming Li and Shijun Liao \footnote{Corresponding author.  Email address: sjliao@sjtu.edu.cn}

\vspace{0.3cm}

State Key Laboratory of Ocean Engineering\\
School of Naval Architecture,Ocean and Civil Engineering\\  Shanghai Jiaotong University, Shanghai 200240, China

 \end{center}

\hspace{-0.6cm}{\bf Abstract}  {
\em A system of two-dimensional,  two  coupled   Faraday   interfacial   waves  is experimentally observed at the two interfaces of the three layers of fluids (air, pure ethanol  and  silicon oil) in a  sealed Hele-Shaw cell with periodic vertical vibration.    The upper and lower Faraday waves coexist: the upper  vibrates vertically, but the crests of the lower one  oscillate horizontally with unchanged wave height and a frequency equal to the half of the forcing one of the vertically vibrating basin,   while  the  troughs of the lower one  always   stay  in  the same place (relative to the basin).   Besides, they are strongly coupled:  the wave height of the lower Faraday wave is either a linear function (in the case of a fixed forcing frequency) or a parabolic function (in the case of a fixed acceleration amplitude) of that of the upper, with the same wave length.   In addition, the upper Faraday wave temporarily loses its smoothness at around  $t=T/4$ and $t=3T/4$, where $T$ denotes the wave period, and thus has fundamental difference from the traditional one.   To the best of our knowledge,   this   system  of   the  two  coupled Faraday waves has never been reported.
}

\vspace{0.3cm}

\hspace{-0.6cm}{\bf Key Words}  Faraday waves; multiple layers of fluids;  experimental observation 
\section{Introduction}

The Faraday waves in a vertically oscillating basin were first discovered by Faraday \cite{Faraday1831} and then analyzed by Benjamin and Ursell \cite{Benjamin1954},   who found that these standing waves  vertically  vibrate  with a frequency equal to half of the forcing one of the basin.  These waves can organize in different forms, such as stripes, squares, hexagons \cite{Binks1997}, and even stars \cite{Rajchenbach2013}.  The Faraday instability in viscous fluids was also experimentally investigated by Bouchgl  and  Aniss \cite{Bouchgl2013}.  Thereafter, the motion of the interface between two fluids with different ratios of density  by means of forcing vertical oscillation became a hot topic.   Some extreme steep interfacial waves which oscillate vertically at the interface of two inviscid  fluids  were numerically simulated and their stability was investigated by Mercer and Roberts \cite{Mercer1992}.   The parametric instability analysis for the interface of two viscous fluids was studied by Kumar and Tuckerman \cite{Kumar1994}, who found that the effect of large viscosity on the wavelength selection is substantial.   The two-dimensional Faraday waves of two inviscid fluids were numerically studied by Wright et al. \cite{Wright2000}, the results were also compared with the fully nonlinear numerical simulation by Takagi and Matusumoto \cite{Takagi2011}.  The  instability of Faraday interfacial waves between two weakly viscous layers in a rectangular domain was studied by Hill \cite{Hill2002}.    The spatiotemporal Fourier spectrum of Faraday waves  on the interface of two liquids in a three-dimensional closed  cell were measured by Kityk \cite{Kityk2005}.   The experimental study on Faraday waves in domains with flexible boundaries is implemented by Pucci et al. \cite{Pucci2011, Pucci2013} in the instability of floating fluid drops.  The  walking and orbiting droplets were observed  on the surface of a  liquid at a sufficiently high acceleration by Couder et al.  \cite{Couder2005}.   The linear Faraday stability of a two-layer  liquid film with a free upper surface was investigated numerically by Potosky and Bestehorn  \cite{Potosky2016}.   The diffuse interface between two miscible liquids subject to vertical vibration was studied by means of experiments and numerical simulation \cite{Amiroudine2012},  and a time-dependent density gradient is established from the moment when the two layers were placed together \cite{Diwakar2015}.   By singular perturbation theory,  the interfacial wave modes in a two-layer liquid-filled cylindrical vessel were found to become more complex,  as the density ratio increases  from the upper to the lower layer \cite{Chang2014}.

In this letter we experimentally investigate the system of the two coupled interfacial Faraday waves at the interfaces of air and two immiscible liquids in a sealed Hele-Shaw cell with periodic vertical vibration.   The upper liquid is pure ethanol (with the density $\rho_1$ =  791kg/m$^3$ and the viscosity $\mu_1$ = 0.0011Pa s) and the lower  is silicon oil (methyl-silicone-I, with the density $\rho_2$ = 970 kg/m$^3$ and the viscosity $\mu_2$ = 0.35 Pa  s).  Above the two immiscible liquids is the air.  So, there exist three layers of different fluids and two interfaces.   One is the interface between the air and the pure ethanol, called the upper interface.   The other is the interface between the pure ethanol and the silicon oil, called the lower interface.

\section{Experimental setup}

The experimental setup is as follows. A  Hele-Shaw cell (made of PMMA) with 300 mm length, 2mm width and 60mm depth is filled with two immiscible fluids: the upper is pure ethanol (4mm in depth) and the lower is silicone oil (8mm in depth).  For the sake of observation convenience,  a very small amount of phenol red is added in pure ethanol. The cell is fixed on a horizontal shaker and guided with a vertical sinusoidal vibration.  The forcing frequency (denoted by $f$) and the acceleration amplitude (denoted by $A$) of the shaker are output by a closed-loop control system.  A high-speed camera is positioned perpendicular to the front of the cell to record the evolution of the upper interface (between the air and pure ethanol) and the lower interface (between the two liquids).  The temperature is nearly $20^{\circ}$C. Considering volatility of pure ethanol, the cell is sealed (i.e. the depth of the air is 48 mm) and the pure ethanol is replaced every twenty minutes.

\begin{figure}
  \centering
  \rotatebox{-90}{\scalebox{0.4}{\includegraphics{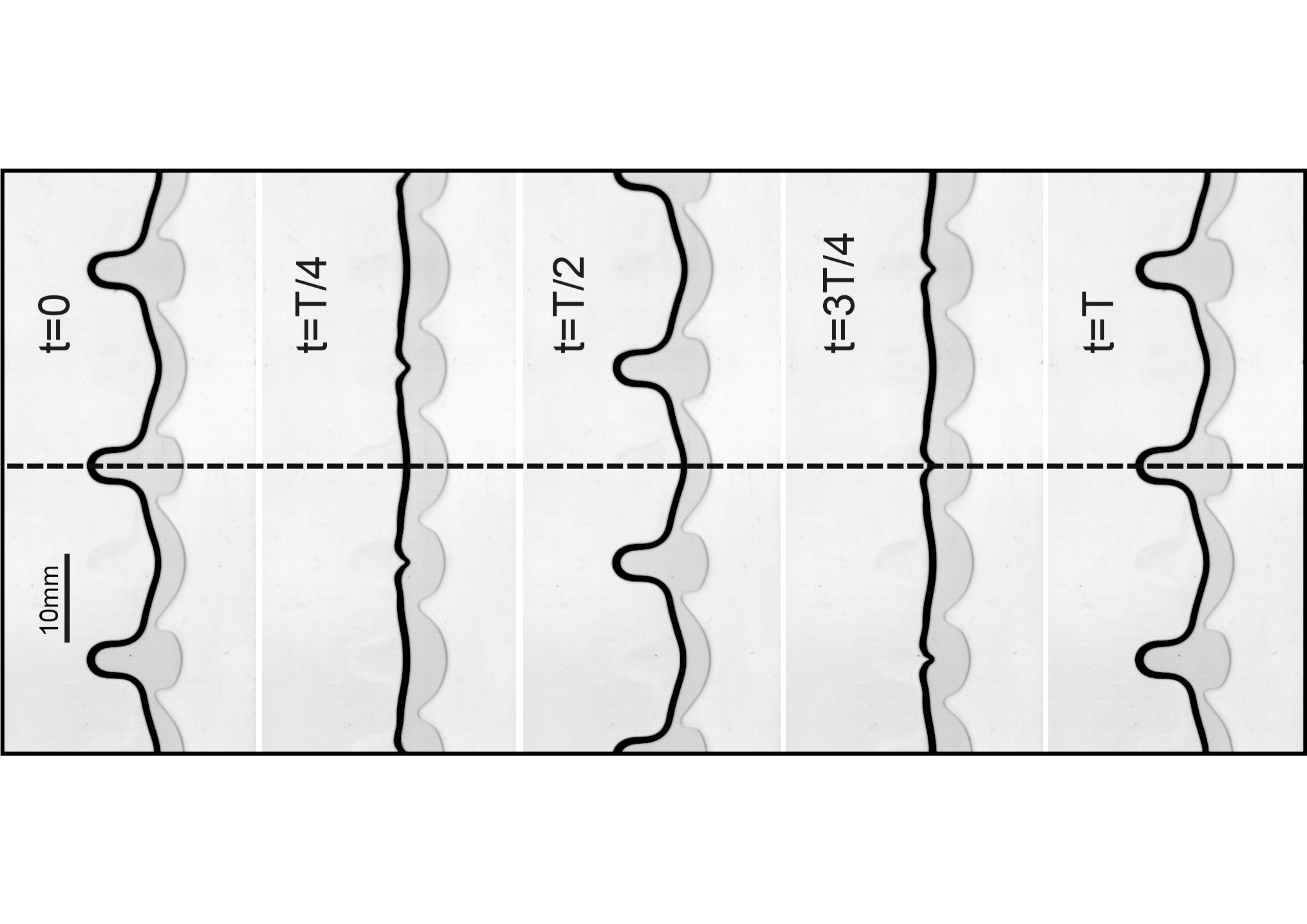}}}
  \caption{ (Colour online) The system of two coupled Faraday waves in the case of the forcing frequency $f $=18 Hz and the acceleration amplitude  $A$ = 17 m/s$^2$, where $T$ denotes the wave period.  For more details, please see the corresponding movie. }
  \label{fig:1}
\end{figure}

\section{Experimental results}

When the Hele-Shaw cell vibrates vertically with the forcing frequency $f$=18 Hz  and the acceleration amplitude $A$=17 m/s$^2$,  we observed a  system of two coupled  Faraday waves.   For details, please see figure~1 and the corresponding movie.  At the upper interface,  there exists a standing wave that oscillates {\em vertically} in a similar way like a traditional Faraday wave (but with a few fundamental differences mentioned later), call the upper Faraday wave.  At the lower interface, there exists a standing wave whose crests  oscillate {\em horizontally} with an unchanged height and a frequency equal to half of the forcing frequency of the {\em vertically} vibrating basin, called the lower Faraday wave. To the best of our knowledge, such kind of horizontally oscillating Faraday waves have never been reported.  The upper and lower Faraday waves coexist and are strongly coupled,  with the same period (denoted by $T$) and the same wave length (denoted by $L$).  At $t=0$,  the crest of the  upper Faraday wave reaches its maximum height, below which there are two adjoining crests of the lower Faraday wave that are in the shortest distance (denoted by $\delta_{min}$). Thereafter,  the crest of the upper Faraday wave falls vertically until it becomes a trough at $t=T/2$, while the above-mentioned two  adjoining crests of the lower Faraday wave depart horizontally from each other with almost unchanged height, and their distance (denoted by $\delta$) increases until it reaches the maximum (denoted by $\delta_{max}$) at $t=T/2$.   As the time further increases, the trough  of  the  upper Faraday wave  moves upwards, but temporarily loses its smoothness at $t=3T/4$, and then becomes a crest again that reaches its maximum at $t=T$.   In the same time,  the two adjoining crests of the lower Faraday wave horizontally approach each  other with the unchanged height until $\delta$ decreases to $\delta_{min}$ at $t=T$.   Note that all troughs of the lower Faraday wave are still (relative to the Hele-Shaw cell) on the same horizontal line, and the distance of any two adjoining troughs is equal to the half of the wave length $L$ of the upper Faraday wave.    However, unlike the upper Faraday wave which has a symmetry about crest,  the lower Faraday wave  loses its symmetry about the crest, although both of the upper and lower Faraday waves  retain the symmetry about the trough.  Besides,  unlike the traditional Faraday wave, the upper Faraday wave temporarily loses its smoothness at around  $t=T/4$ and $t=3T/4$.  It implies that the upper and lower Faraday waves strongly interact each other.  In addition, it is found that,  using the same forcing frequency $f$ = 18 Hz and the same acceleration amplitude $A$=17  m/s$^2$,  we can {\em not} observe any Faraday waves if there exists only the 8mm silicone oil (with the air) in the sealed Hele-Shaw cell, or if we increase the depth of  pure ethanol up to 10mm.   This  phenomenon strongly suggests that the lower horizontally oscillating Faraday wave is excited by the upper vertically vibrating Faraday wave via the viscous friction on the interface between the two immiscible liquids.

\begin{figure}
  \centering
  \rotatebox{0}{\scalebox{0.29}{\includegraphics{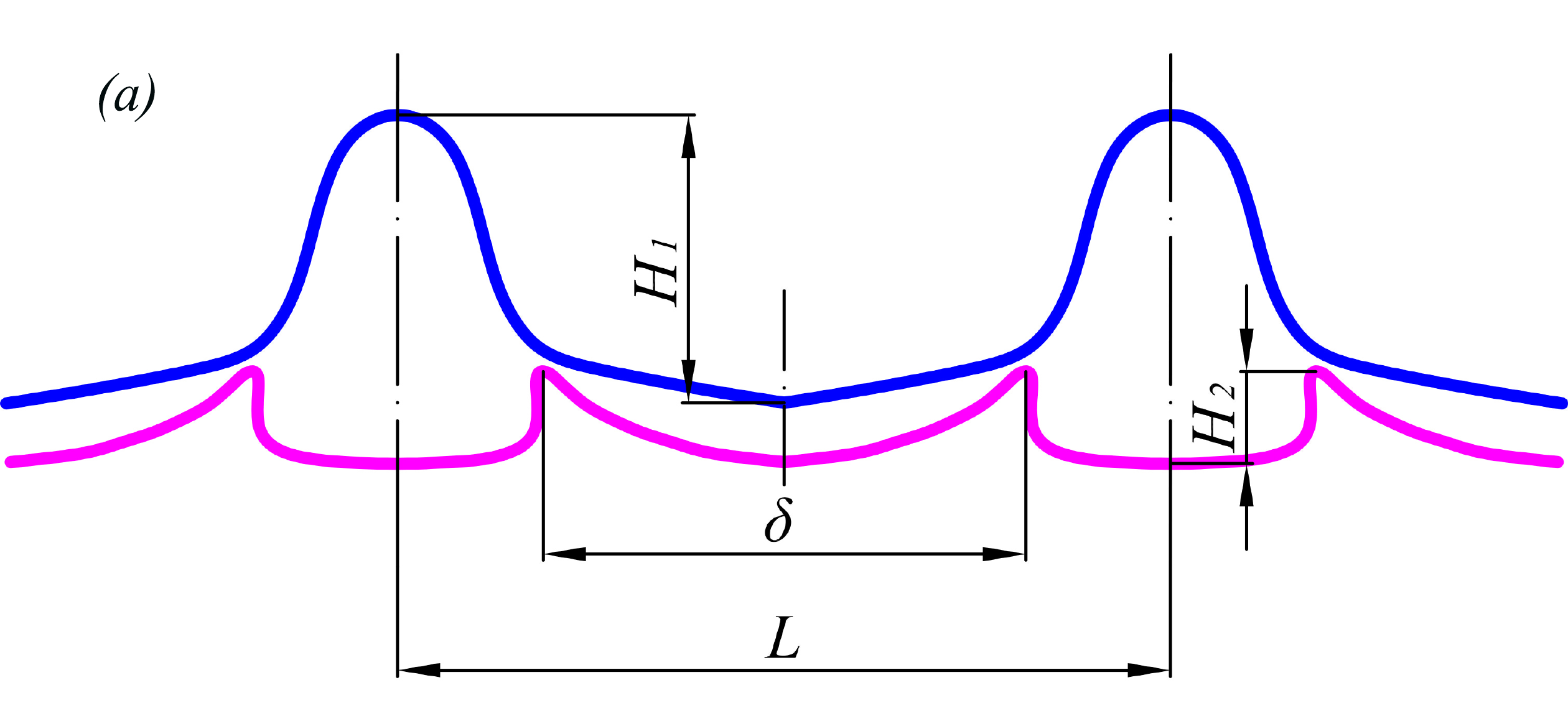}}}
  \rotatebox{0}{\scalebox{0.1}{\includegraphics{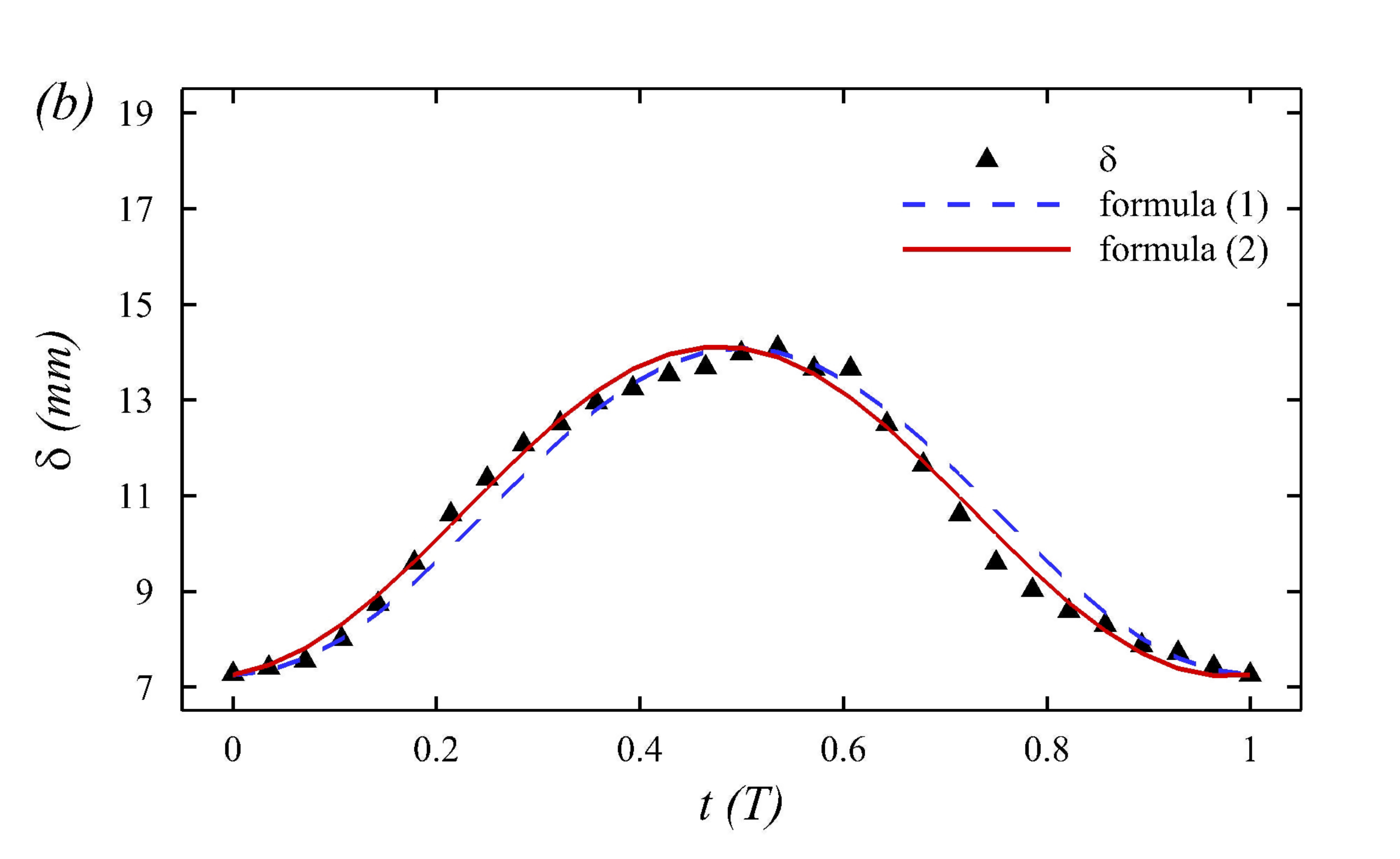}}}
  \caption{ (Colour online) (a) Schematic illustration of the upper and lower Faraday waves.  $H_1$ and $H_2$ denote the wave height of the upper and lower Faraday waves,   $L$ is their wave length, $\delta$ is the distance of the two crests of the lower Faraday wave, respectively. (b) Variation of $\delta(t)$ in case of $f$ =18 Hz and  $A$ = 17 m/s$^2$. }
\label{fig:2}
\end{figure}

The schematic illustration is as shown in figure~2(a), where $H_1$ and $H_2$ denote the wave height of the upper and lower Faraday waves, $L$ denotes their wave length, $\delta$ is the distance between the two adjoining crests of the lower Faraday wave, respectively.   In the case of $f$ =18 Hz and $A$ = 17 m/s$^2$, the time-dependent  variation  of $\delta$ is as shown in figure~2(b), which can be fitted by a simple formula
\begin{equation}
\delta(t) = \frac{1}{2}\left(\delta_{max}+\delta_{min} \right) -\frac{1}{2}\left(\delta_{max}-\delta_{min} \right) \; \cos(\pi f t)   \label{delta:theory}
\end{equation}
with $\delta_{max} = 14.08$ mm and $\delta_{min} = 7.26$ mm  in a good agreement with the measured data,  and by the fitted formula
\begin{equation}
\delta(t) = \frac{1}{2}\left(\delta_{max}+\delta_{min} \right)-\frac{1}{2}\left(\delta_{max}-\delta_{min} \right) \; \cos(\pi f t)\\
 + B \sin(\pi f t) \label{delta:fit}
\end{equation}
with $B=0.4865$ mm  in a better agreement.

\begin{figure}
  \centering
  \rotatebox{0}{\scalebox{0.1}{\includegraphics{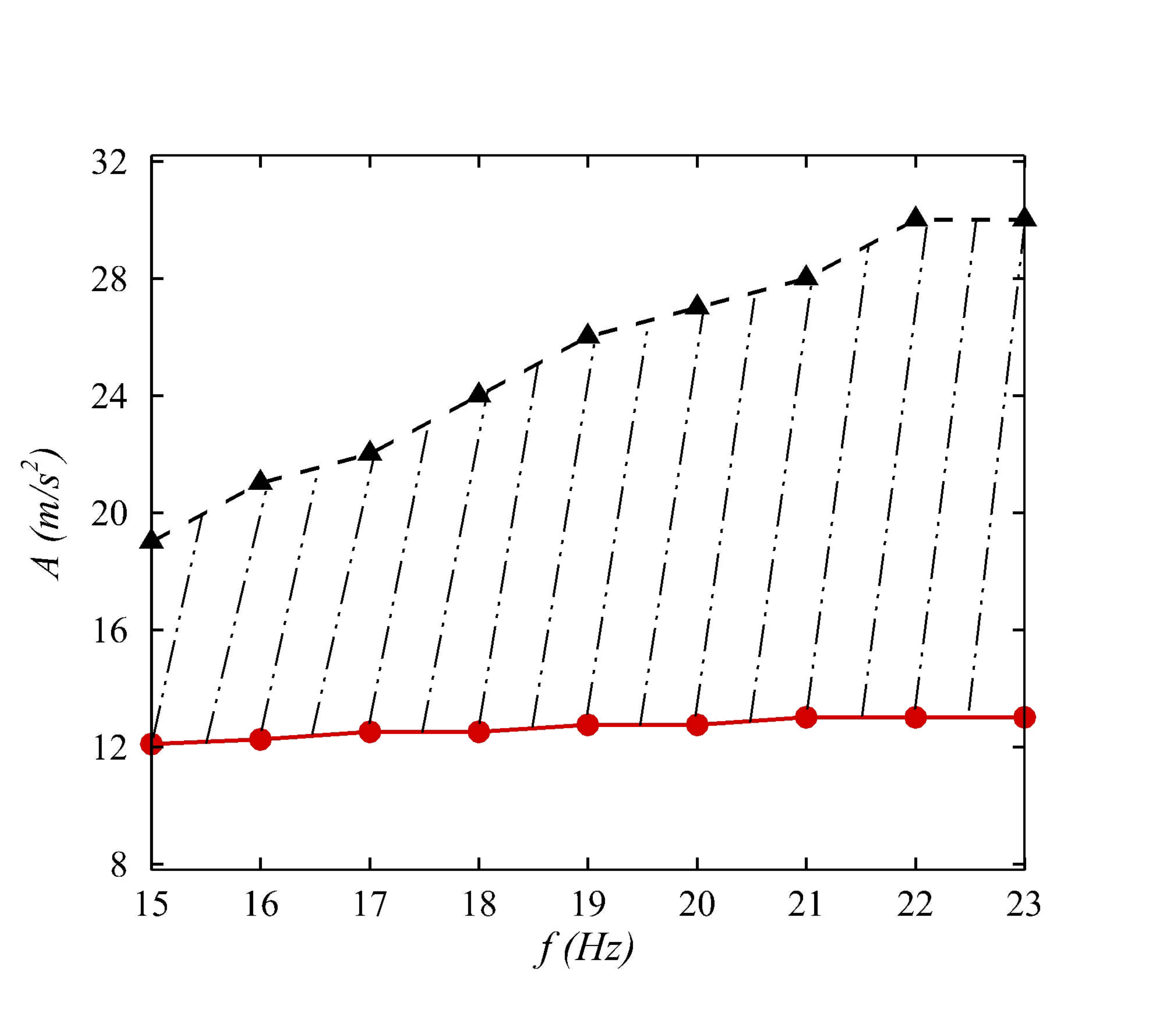}}}
  \caption{ (Colour online) Existence window of $A$ versus $f$ for the couple two Faraday waves. Solid line:  the lower threshold; Dashed line: the upper threshold.}
\label{fig:3}
\end{figure}

With $f$ fixed at 18Hz, the system of the two coupled Faraday waves can be observed within a region of the acceleration amplitude 12.5 m/s$^2$  $\leq A\leq $ 24  m/s$^2$. When $A < $ 12.5 m/s$^2$,  no interfacial waves were observed at all.  When $A>24$ m/s$^2$, the interfacial waves at the upper and lower interface are disordered.  In the cases of $f$ = 23 Hz and $f=$15 Hz, such kind of two coupled Faraday waves are always observed, but with different upper and lower thresholds of $A$.  It is found that  there exists the corresponding upper and lower thresholds of $A$ for a given  forcing frequency $f$, as shown in figure~3.   Note that the lower threshold of $A$  increases with the frequency $f$ very slowly, but the upper threshold rises rapidly.

\begin{figure}
\centering
  \rotatebox{0}{\scalebox{0.1}{\includegraphics{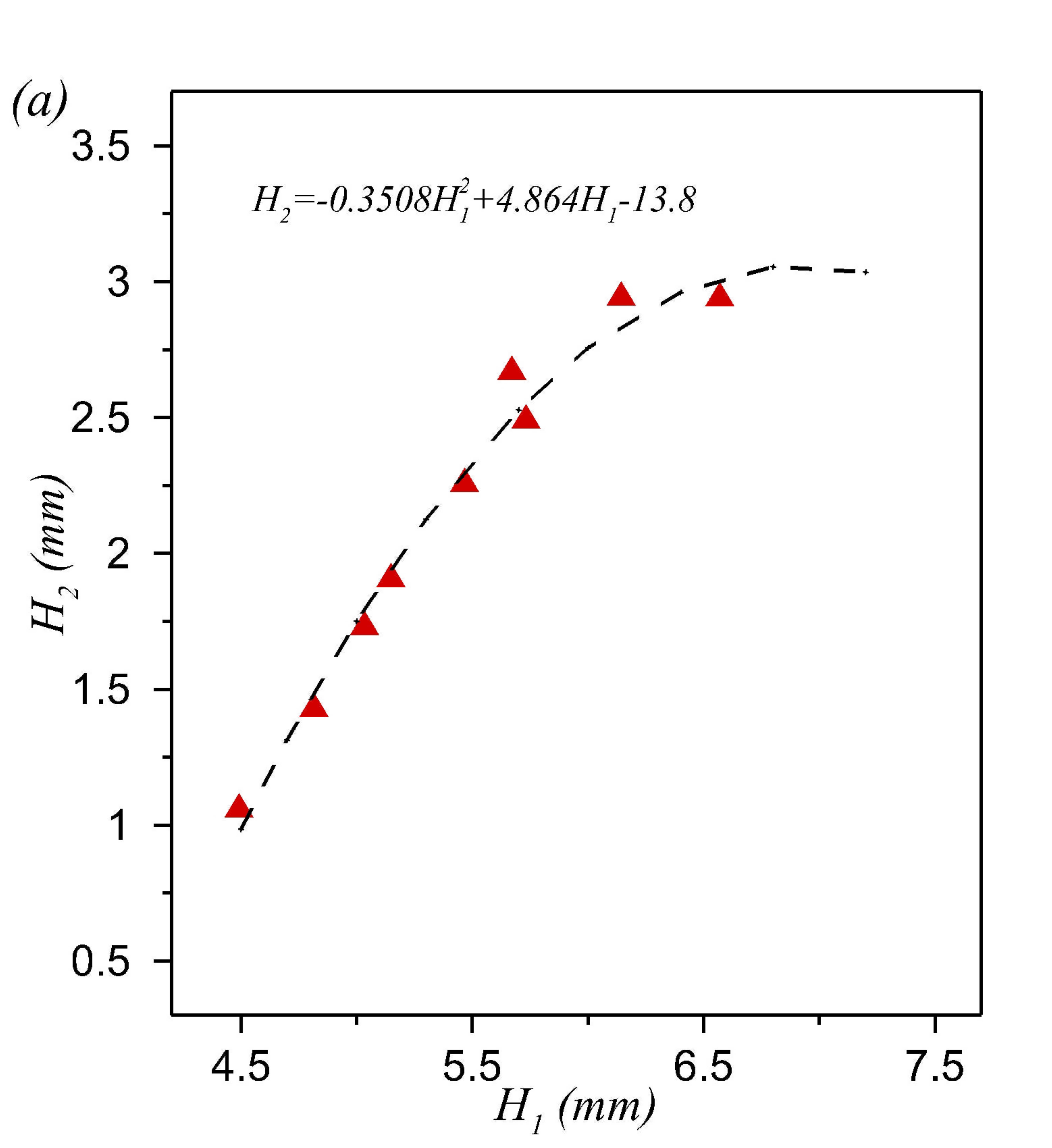}}}
  \rotatebox{0}{\scalebox{0.1}{\includegraphics{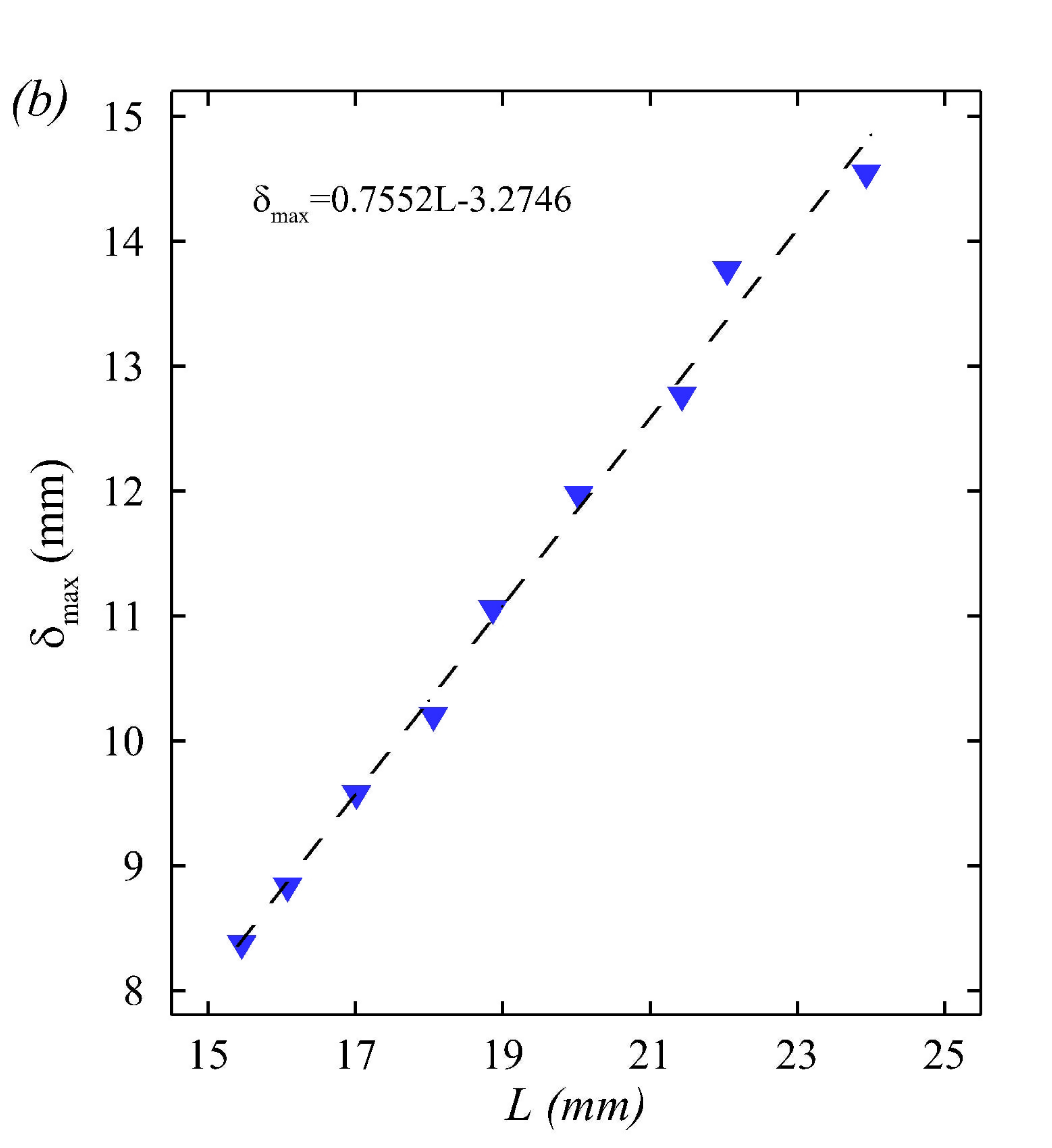}}}
  \rotatebox{0}{\scalebox{0.1}{\includegraphics{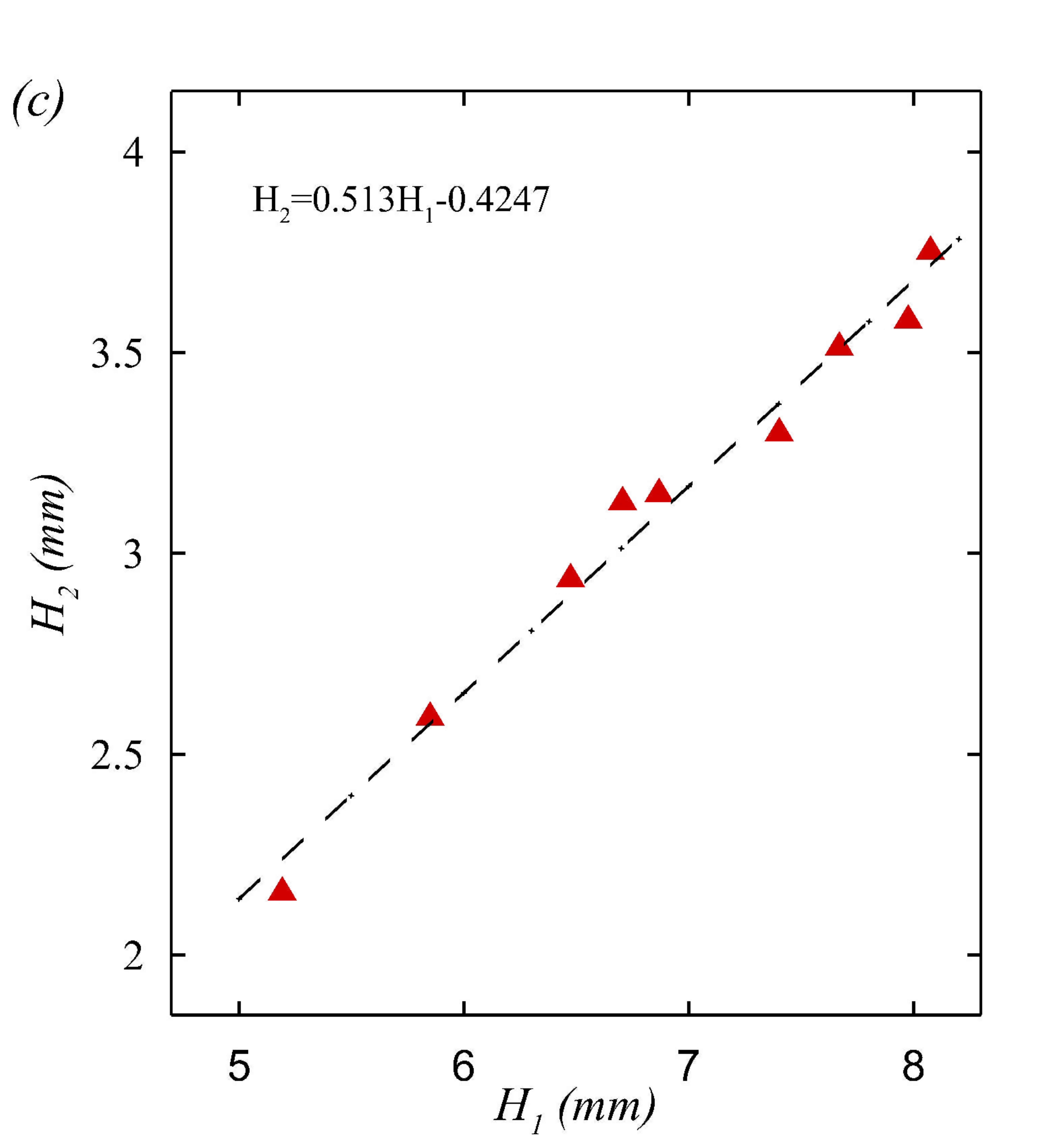}}}
  \rotatebox{0}{\scalebox{0.1}{\includegraphics{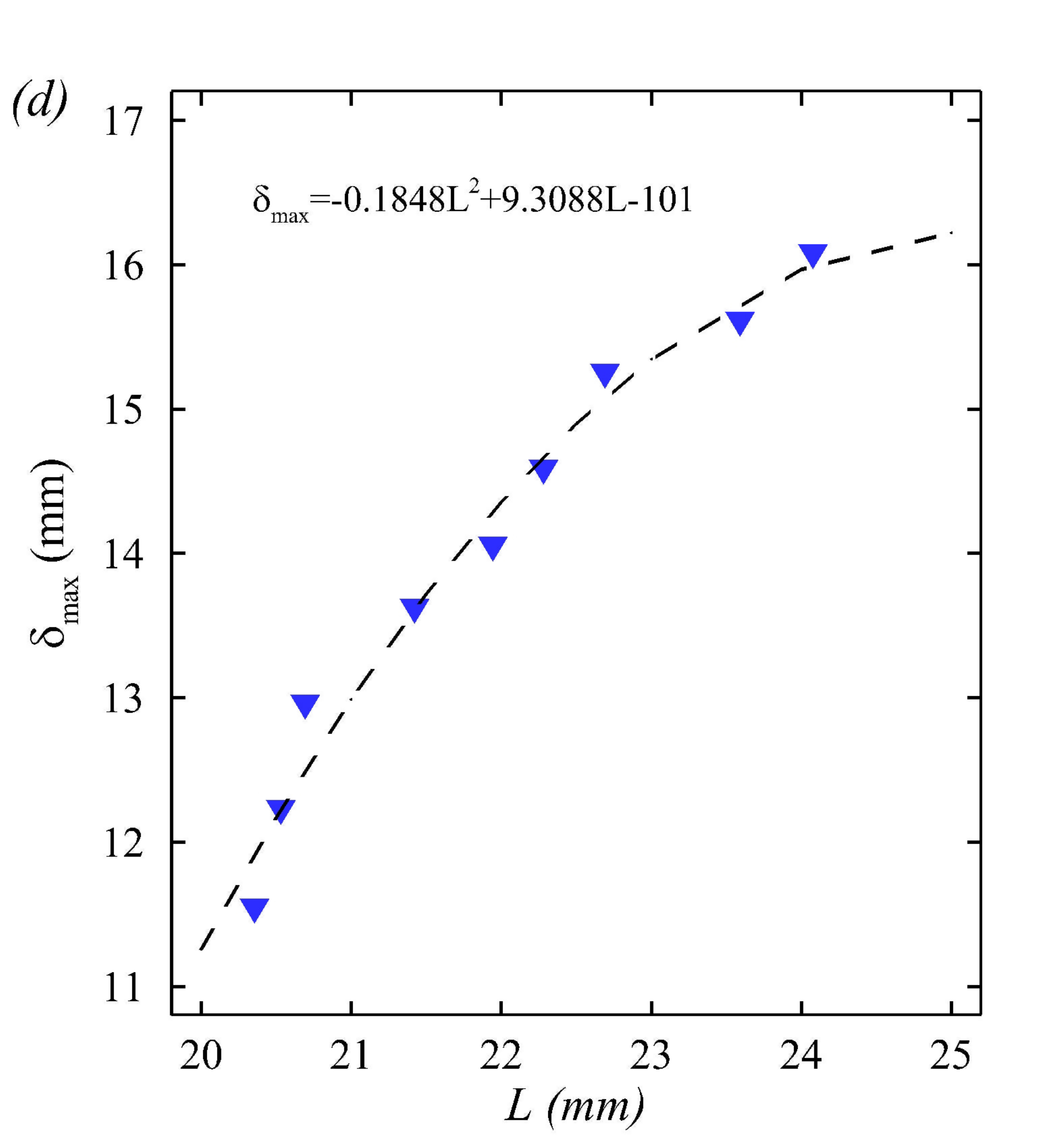}}}
  \caption{ (Colour online) Relations between $H_1$ and $H_2$, $L$ and $\delta_{max}$ of the coupled two Faraday waves.  (a) and (b):  in case of  $A$=15 m/s$^2$ and  15 Hz $\leq f \leq 23 $ Hz;   (c) and (d):  in the case of $f$=18 Hz and  14 m/s$^2$ $\leq A\leq$ 22 m/s$^2$; Dashed-line:  the fitting formulas.}
\label{fig:4}
\end{figure}

In the case of the fixed acceleration amplitude $A$= 15 m/s$^2$  with the different forcing frequency  $f$ in the region of 15 Hz $\leq f \leq 23$ Hz, the wave height $H_2$ of the lower Faraday wave is a parabolic function of the wave height $H_1$ of the upper one, but $\delta_{max}$  has a linear relationship with the wave length $L$,  respectively, as shown in figure~4(a,b).   In the case of  the fixed forcing frequency  $f$=18 Hz  with  the different acceleration amplitude $A$ in the region of  14 m/s$^2$ $\leq A\leq$ 22 m/s$^2$,   the wave heights $H_2$ of the lower Faraday wave has a linear relationship with the wave height $H_1$ of the upper one, but $\delta_{max}$ is a parabolic function of the wave length $L$, respectively,   as shown in figure~4(c,d).  These phenomena reveal the close relationship and strong coupling between the upper and lower Faraday waves.

\begin{figure}
\centering
  \rotatebox{0}{\scalebox{0.1}{\includegraphics{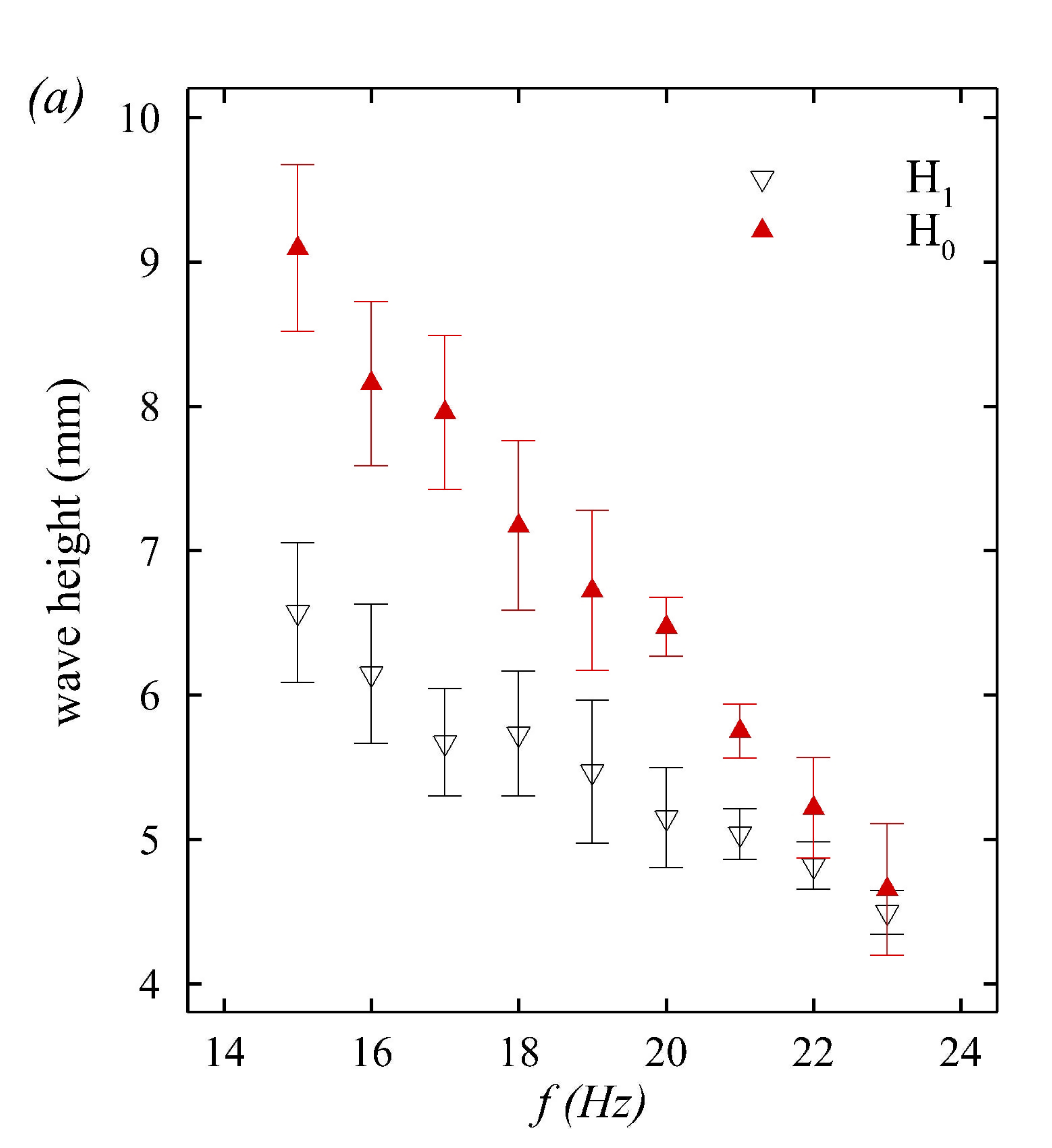}}}
  \rotatebox{0}{\scalebox{0.1}{\includegraphics{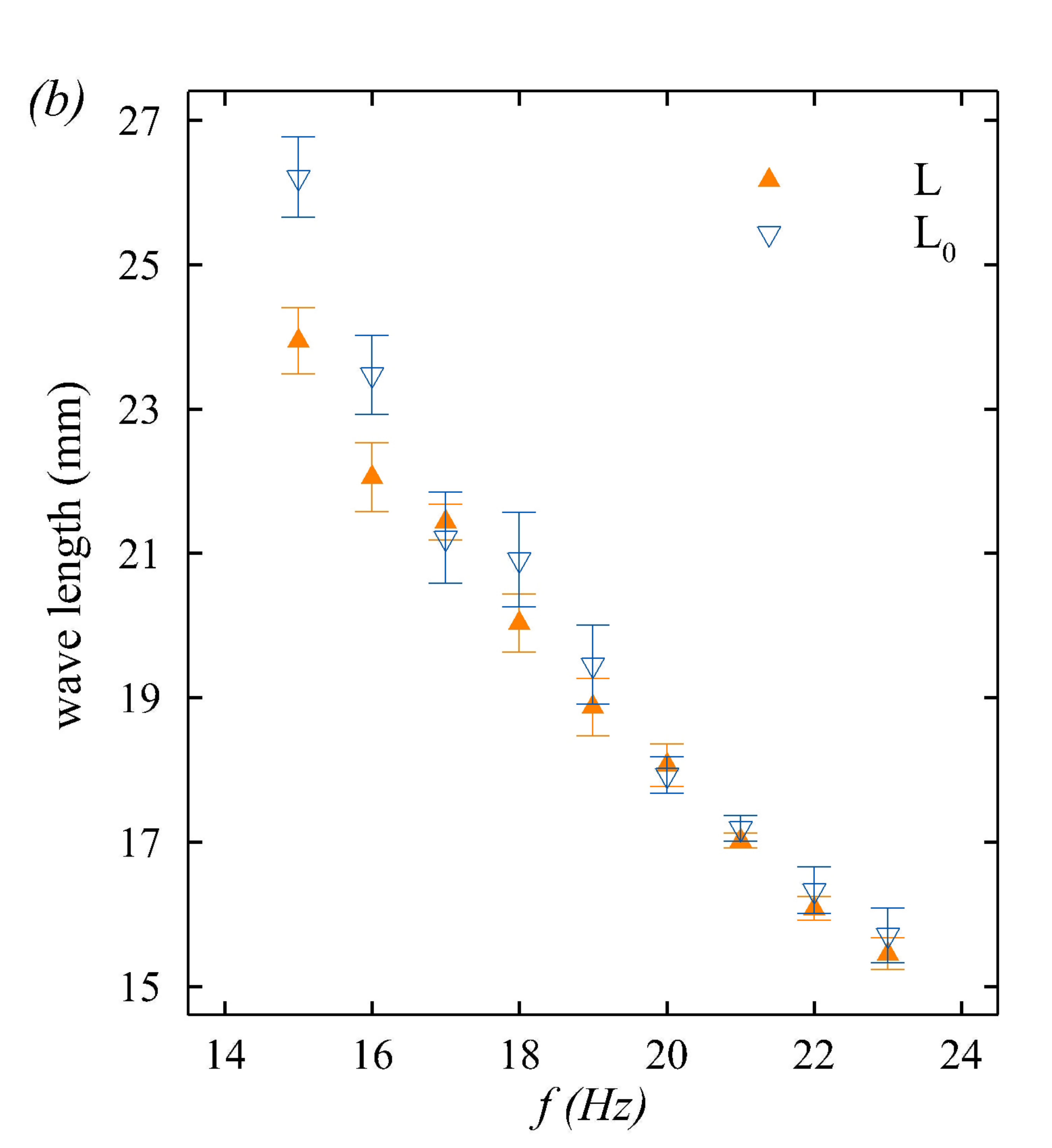}}}
  \rotatebox{0}{\scalebox{0.1}{\includegraphics{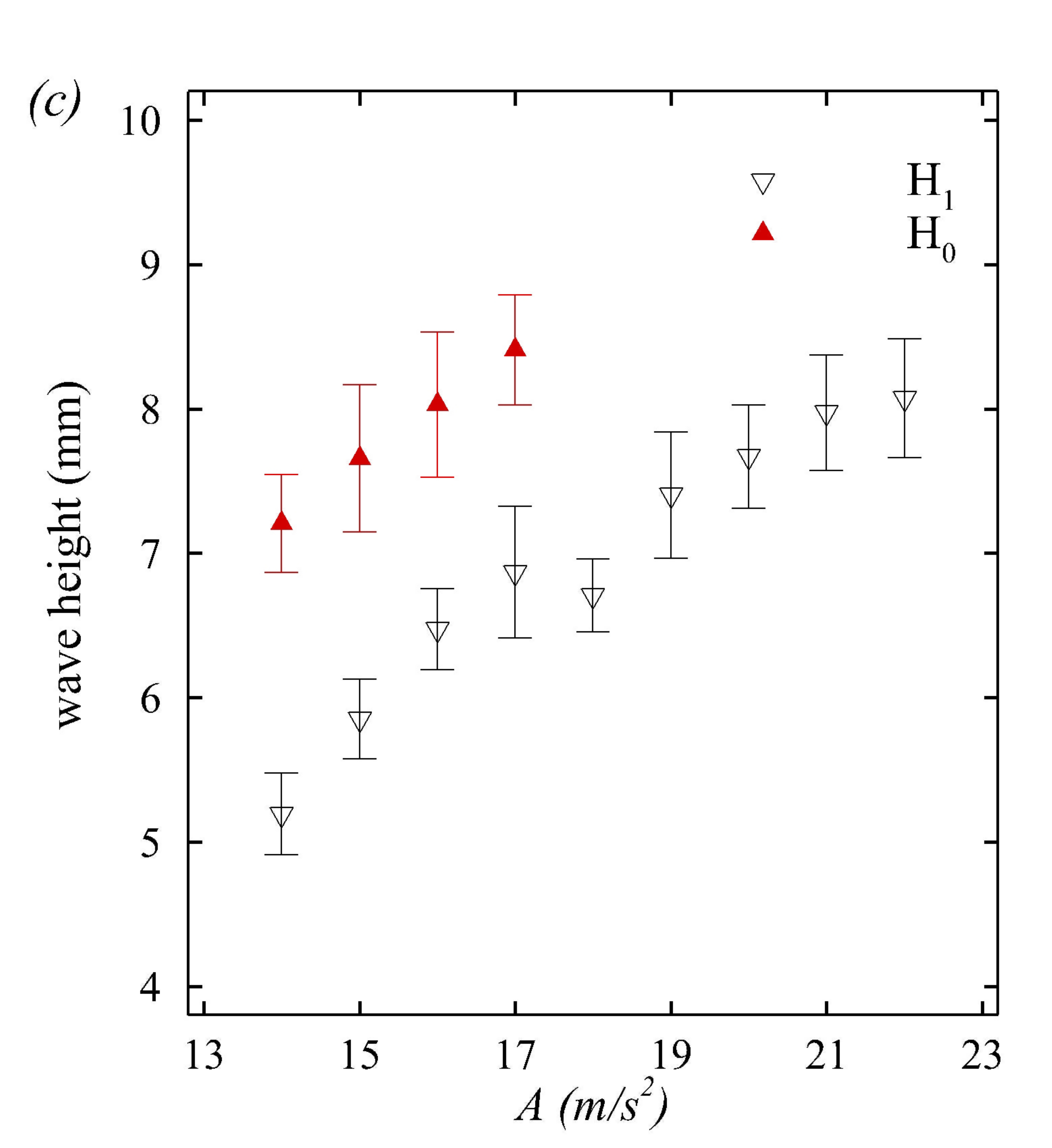}}}
  \rotatebox{0}{\scalebox{0.1}{\includegraphics{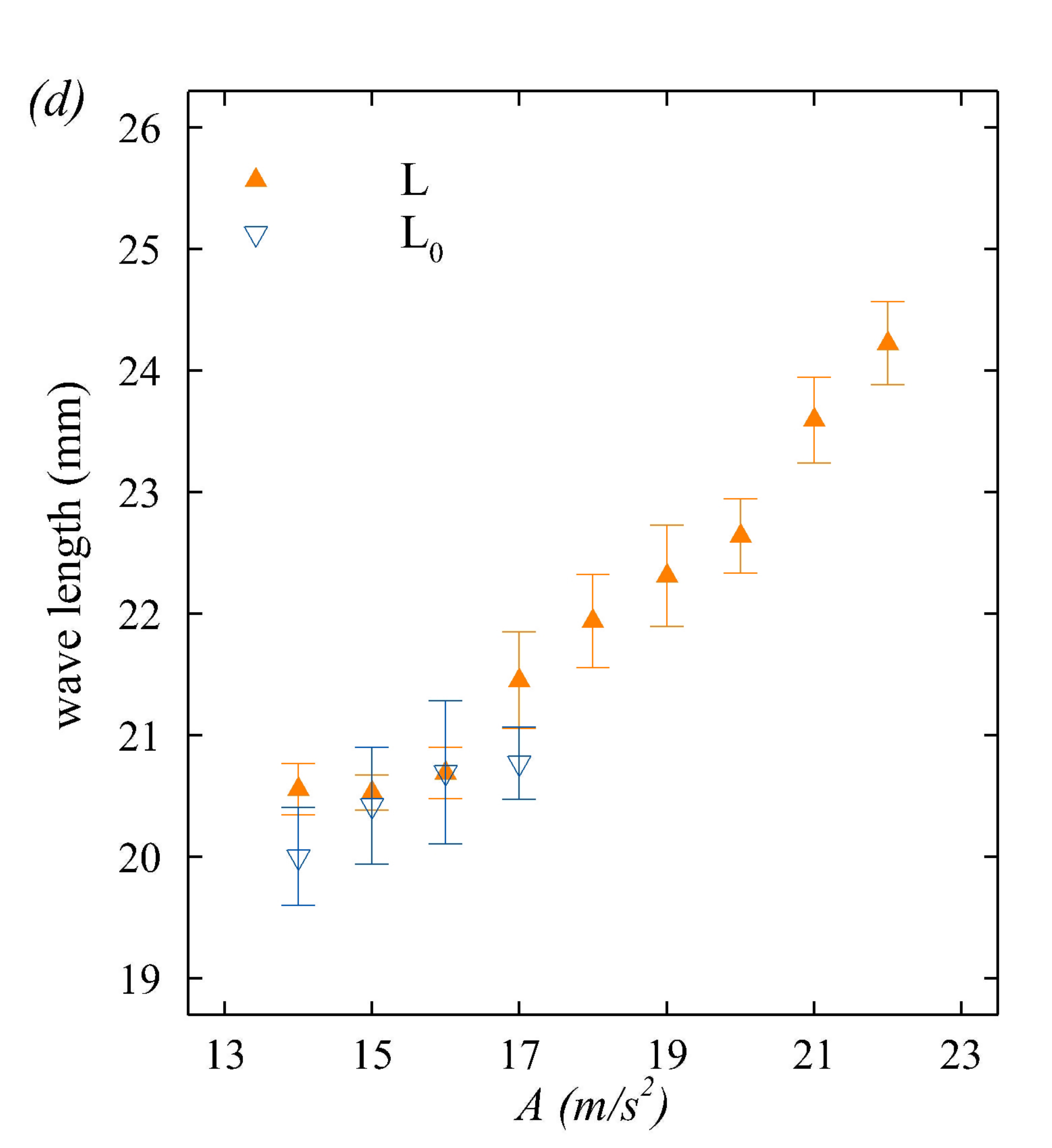}}}
  \caption{(Colour online) Comparison of the upper Faraday wave and the traditional Faraday wave of pure ethanol with the same physical parameters (but without the layer of  silicon oil below), where $H$ and $L$ denote wave height and wave length of the upper Faraday wave, $H_0$ and $L_0$ denote those of the traditional one.    (a) and (b):  in case of $A$=15 m/s$^2$ and 15 Hz $\leq f\leq $ 23 Hz;   (c) and (d):  in the case of $f$=18 Hz and 14 m/s$^2$ $\leq A\leq$ 22 m/s$^2$.}
\label{fig:5}
\end{figure}

Note that the upper Faraday wave looks like the traditional Faraday wave at the interface of two immiscible fluids only (such as water and air).      For the sake of comparison, we measured the traditional Faraday wave at the interface  of the air and pure ethanol with the same depth of 4mm  (but {\em without} silicon oil)  in the same Hele-Shaw cell \cite{Li2015, Li2016}.    In the case of  the  fixed acceleration amplitude  $A$=15 m/s$^2$ with the different  forcing frequency $f$ in the region of 15 Hz $\leq f\leq $ 23 Hz,  both of the wave height $H_1$ and wave length $L$ of the upper Faraday wave decreases with the increase of the frequency $f$, as shown in figure~5(a,b).   Besides,   it is found that the wave height $H_1$ of the upper Faraday wave is always smaller than the wave height $H_0$ of the traditional one,  but the wave length  $L$ of the upper Faraday wave is almost the same as that of the traditional one, respectively.   This is easy to understand, since the upper layer liquid (pure ethanol) transfers some  kinetic energy to the lower one (silicon oil) via viscous friction at their interface.  In the case of  the fixed frequency  $f$=18 Hz with the different acceleration amplitude $A$ in the region 14  m/s$^2$ $\leq A\leq$ 22 m/s$^2$,  both of the wave height $H_1$ and wave length $L$ of the upper Faraday wave increases with the increase of $A$, as shown in figure~5(c,d).  However,  it  is interesting that the traditional Faraday wave does not exist when $A>$ 17 m/s$^2$, but the system of the  two  coupled  Faraday waves  exists even for the acceleration amplitude $A$ up to 22 m/s$^2$.   It means that, for a given forcing frequency $f$,  the upper threshold of $A$ for the coexistence of the  two  coupled  Faraday waves is larger than that for only one traditional Faraday wave.   It indicates that the system of the two coupled  Faraday waves is stable even for a large acceleration amplitude  $A$ that corresponds to a high nonlinearity.     Note that, for a given frequency $f$,  more kinetic energy is needed to excite the system of the two coupled Faraday waves than the  only  one traditional  Faraday wave.  In addition,  the wave length $L$ of the upper Faraday wave is almost the same as $L_0$ of the traditional one, although its wave height $H_1$ is always smaller than $H_0$.     Furthermore,  unlike the traditional Faraday wave,  the upper Faraday wave temporarily loses its smoothness  at around $t=T/4$  and  $3T/4$, as shown in figure~1.  All of these indicate that the upper Faraday wave is fundamentally different from the traditional one, although both of them are vertically  vibrating  waves.   Finally, it should be emphasized that the upper and lower Faraday waves coexist and are strongly coupled: neither of them can exist along.

Note that Potosky and Bestehorn \cite{Potosky2016} numerically investigated the linear instability of Faraday waves of the three-layer fluids (air and two immiscible liquids) in a {\em three-dimensional} domain.   However, they only gained the coupled two Faraday waves that vibrate vertically.   Unlike their theoretical investigation,  our physical experiments are related to the {\em two-dimensional} Faraday waves by means of physical parameters   quite  different  from  theirs:   the ratio of the viscosity $\mu_2/\mu_1\approx 318.2$ in our experiment  is  about  22 times larger than that considered by Potosky and Bestehorn \cite{Potosky2016}.

\section{Conclusions}

In conclusion, we  experimentally  observed a system of the two-dimensional, two coupled  Faraday waves at two interfaces of three layers of fluids (air, pure ethanol and  silicon oil)  in  a   sealed  Hele-Shaw cell with periodic vertical vibration.   The upper Faraday wave vibrates vertically, and the lower oscillates horizontally.    They coexist and are strongly coupled.   This  system of two coupled Faraday waves has never been reported, to the best of our knowledge.   So, it fleshes out the picture of Faraday waves as a type of vertical standing waves.   They also bring us  some new challenges in theoretical analysis  and  numerical simulations.

\section*{Acknowledgement}  This work is partly supported by National Natural Science Foundation of China (Approval No. 11272209 and 11432009).

\section*{References}
\bibliographystyle{elsarticle-num}
\bibliography{experiment}
\end{document}